\documentclass[, 5p]{elsarticle}
\usepackage[utf8]{inputenc}
\usepackage[T1]{fontenc}
\usepackage{siunitx}
\usepackage{amsmath}
\usepackage{amssymb}
\journal{Optics Communications}

\begin{document}

\begin{frontmatter}
\title{Local Optimization of Wave-fronts for optimal sensitivity PHase Imaging (LowPhi)}

\author[1,2,3]{Thomas Juffmann} \corref{cor1}
\ead{thomas.juffmann@univie.ac.at}
\author[1]{Andr\'es de los R\'ios Sommer}
\author[1]{Sylvain Gigan}
\cortext[cor1]{Corresponding author} 

\address[1]{Laboratoire Kastler Brossel, Sorbonne Universit\'e, Ecole Normale Superi\`eure, Coll\`ege de France, CNRS UMR 8552, 24 rue Lhomond, 75005 Paris, France}
\address[2]{Faculty of Physics, University of Vienna, Boltzmanngasse 5, A-1090 Vienna, Austria}
\address[3]{Department of Structural and Computational Biology, Max Perutz Laboratories, University of Vienna, Campus Vienna Biocenter 5, A-1030 Vienna, Austria}

%\affil[+]{these authors contributed equally to this work}

%\keywords{Keyword1, Keyword2, Keyword3}

\begin{abstract}
Phase contrast microscopy is an invaluable tool in the biosciences and in clinical diagnostics. However, its sensitivity varies significantly across a given sample because it is optimized for one specific mean value of the sample induced phase shifts. Here, we demonstrate a technique based on wavefront shaping that optimizes the sensitivity across the field of view and for arbitrary phase objects. We show significant sensitivity enhancements, both for engineered test samples and red blood cells. Our technique adds on naturally to commercial phase contrast microscopes, reduces imaging artifacts, and, once initialized, allows for quantitative single-frame phase microscopy.
\end{abstract}


\end{frontmatter}

%\flushbottom
%\maketitle
%\thispagestyle{empty}

\section*{Introduction}

Interferometric microscopy techniques are invaluable for the study of biological specimens such as cells, and have found both scientific and diagnostic applications \cite{Lee2013, Park2018}. In these techniques, a signal wave $E_S=|E_S | e^{i\phi}$, is interfered with a reference wave $E_R=|E_R | e^{i\alpha}$. Interference yields an intensity $I=|E_S |^2+|E_R |^2 + 2|E_R ||E_S |\cos{(\phi-\alpha)}$, which allows retrieving information about the signal phase $\phi$, with a sensitivity $\frac{dI}{d\phi}$ proportional to $\sin{(\phi-\alpha)}$. The sensitivity is best for:  
\begin{equation}
\phi-\alpha=(2n-1)\pi/2
\label{eq:eq00}
\end{equation} 
, where $\mid n \mid =1,2,3$,…. 
For flat samples, interferometric techniques allow for the detection of atomic steps on surfaces \cite{Tolansky1951}, and for the unlabeled detection of single proteins \cite{Piliarik2014}. 
However, for various samples, such as cells, $\phi$ varies spatially over the full range $(-\pi, \pi)$. Since $\alpha$ is often fixed (e.g. in phase contrast microscopy \cite{Zernike1942}), this leads to a sensitivity that goes to zero for certain regions of the sample. This can be avoided stepping the phase of the reference beam \cite{Popescu2004, Wang2011} or by varying the angle of the incoming illumination \cite{Tian2015}. While being applied very successfully, such approaches naturally increase acquisition times. Quantitative single shot techniques have been demonstrated, relying either on off-axis holography \cite{Wang2010} necessitating a trade-off in spatial resolution, on parallelization using beam splitters \cite{Zuo2013} increasing experimental complexity, on wavefront sensors \cite{Bon2009} or on polarization encoding \cite{Chu2012}, both of which are only sensitive to phase gradients. Either way, varying $\alpha$ means that the total sensitivity of the combined acquisitions will be lower than the sensitivity that would be reached, if $\alpha$ was chosen according to equation \eqref{eq:eq00} for all regions of the specimen. 

Wave-front shaping, the ability to digitally control the phase of incident light across large pixel arrays using MEMS or Liquid Crystal based technologies, has recently emerged as a powerful tool in microscopy and imaging. Various beam shaping techniques have been developed to enhance the capabilities of phase microscopes \cite{Ng2004,Maurer2011,Bernet2006, Nguyen2017}, mostly by tailoring phase masks in the diffraction plane. Recently, compressive phase imaging has been realized using a digital mirror device in an image plane \cite{Soldevila2018}.  

Here, we enhance phase microscopy by shaping a wave-front in a plane that is conjugate to the sample plane. We demonstrate how optimal sensitivity across the whole field of view can be obtained even if the specimen induced phase changes $\phi(x,y)$ cover the full range $(-\pi, \pi)$. We demonstrate the principle of our new technique in a standard phase contrast configuration, a common-path technique that is intrinsically stable and widely used. We stress however, that LowPhi can also be applied to other phase imaging techniques, like holography, in which either the reference or the signal beam can be appropriately shaped. 

Phase contrast microscopy \cite{Zernike1942} is based on the insight that the optical field after sample interaction can be written as the sum of a plane (unscattered) reference wave and a signal (scattered) wave $E=|E| e^{i\psi(x,y)}=|E_R |+|E_S (x,y)| e^{i\phi(x,y)}$. In the Fourier plane (after propagation through a 2f setup), the reference wave is focused onto a point, while the scattered wave is spread out across the plane, which allows adding a phase $\alpha$ to the reference wave locally, without affecting the scattered wave: 
\begin{equation}
E_F=|E_R |e^{i\alpha}+|E_S (x,y)| e^{i\phi(x,y)}
\label{eq:eq0}
\end{equation}
Propagation through another 2f setup yields an image of intensity
\begin{equation}
I(x,y)=|E_S(x,y)|^2+|E_R|^2+\\2|E_S(x,y)||E_R|\cos{(\phi(x,y)-\alpha)}
\label{eq:eq1}
\end{equation}
For $\alpha=\pm\pi/2$, the conventional schemes of positive and negative phase contrast are retrieved, which represent optimal choices only if $\phi(x,y)\approx 0$. 

Our technique (Fig. \ref{fig:Fig1}a and b) uses a spatial light modulator (SLM) to ensure that this condition is met everywhere across the field of view of the image. First, a phase image of the sample is taken using a traditional quantitative phase imaging technique \cite{Popescu2004}. Then, an SLM in a plane conjugate to the specimen plane (S) is used to subtract the measured phase-shifts from the wave-front. A consecutive quantitative phase image now yields the error of the measurement $\psi(x,y)-\psi_{meas} (x,y)$, which can be due to statistical or systematic errors (e.g. due to imaging artifacts such as halo effects commonly observed in phase microscopy). If the error is non-negligible, the procedure can be applied iteratively until $\psi(x,y)-\psi_{meas}(x,y)\ll1$ across the field of view. 
The choice $\alpha=\pm \pi/2$ now yields optimal sensitivity and quantitative phase measurements (see analysis section) across the image. For dynamic studies this means that initialization of the SLM at time $t_0$, will allow for optimal sensitivity at all following times $t_1>t_0$, as long as the temporal changes of the phase distribution are $\lesssim 0.5$. Even if they become significant, LowPhi can still operate quantitatively, as long as the phase mask displayed on the SLM can be updated at a rate sufficient to follow the large scale slow variations of the sample over time. 

\section*{Results}
\subsection*{Setup}
\begin{figure}[!ht]%[htbp]
\centering
\fbox{\includegraphics[width=84mm]{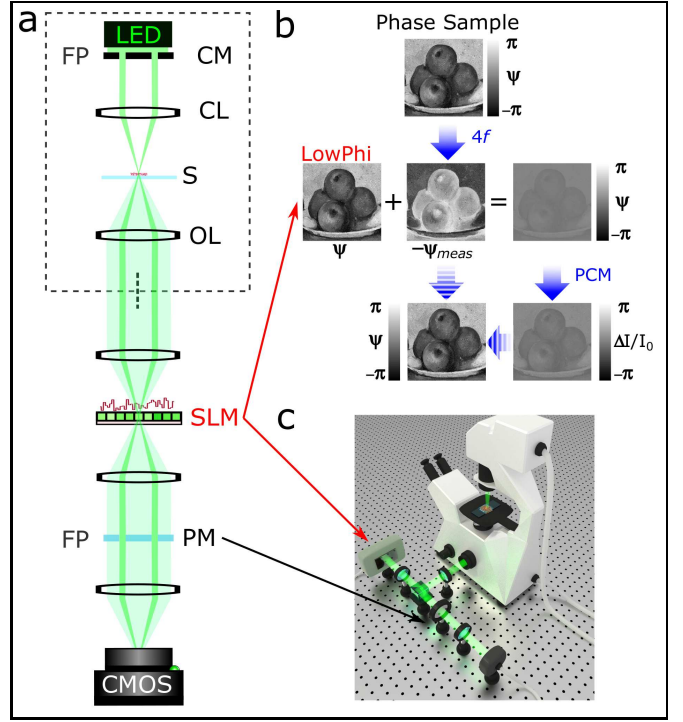}}
\caption{(a and b) Sketch and concept of LowPhi based on a commercial inverted microscope (dashed box with condenser mask (CM), condenser lens (CL), sample (S), and objective lens (OL)). A 4f setup images the wave-front onto a spatial light modulator (SLM). The SLM is used to add the negative of a coarse estimate of the phase distribution to the wave-front. The remaining deviations from a plane wave are given by the error in the coarse estimate. They are imaged using a standard phase contrast microscope (PCM), having a $\pm\pi/2$ phase mask (PM) in a plane conjugate (Fourier plane...FP) to the matching condenser mask (CM). For small errors this transforms the phase distribution into a measurable intensity distribution in a quantitative and linear way (see analysis section), allowing for a straightforward reconstruction (dashed) of the original phase distribution. (c) The setup can be realized as a simple add-on to a commercial (inverted) microscope.}
\label{fig:Fig1}
\end{figure}

The simplified setup is sketched in Fig. \ref{fig:Fig1}. 
The sample (S) is illuminated with light from an LED, and imaged onto a spatial light modulator (SLM, Hamamatsu X10468-04, \SI{20}{\micro\metre} pixel pitch) using a $M=50$x ($NA = 0.75$) objective, which was chosen such that each diffraction limited area of the sample can be properly addressed using the SLM.
After reflection from the SLM, a 4f lens configuration with unity magnification is used to image the wave onto a camera (CMOS).
To perform phase contrast microscopy, a condenser mask (CM) and a matched phase mask (PM) have to be inserted in Fourier planes (FP) before the sample and after the SLM.
These allow for selective application of $\alpha$ to the unscattered wave. While these masks are typically ring-shaped, we opted here for a more versatile solution: a random (but known) pattern, which allows reducing imaging artifacts \cite{Maurer2008}. The required CM was laser printed, the PM was realized using an SLM. In our setup, a folded configuration allowed us to use two regions of a single, reflective SLM to realize both the SLM and the PM patterns (see Appendix). Crucially, LowPhi is simply an add-on to a commercial inverted microscope (see Fig. \ref{fig:Fig1} c). 
\subsection*{LowPhi}
\begin{figure}[t]
\centering
\fbox{\includegraphics[width=84mm]{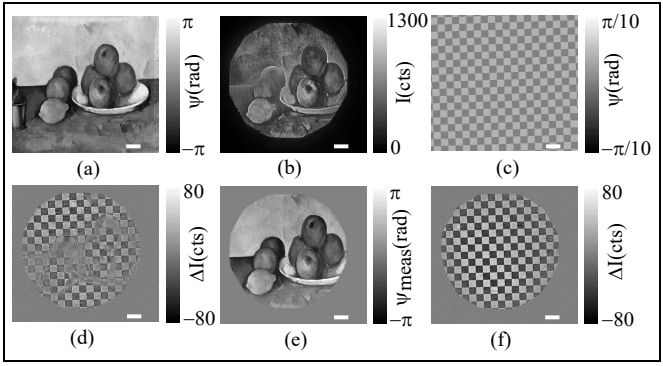}}
\caption{(a) Cezanne’s “Still Life with Apples” converted into a phase distribution ($\psi_{CZ}$). (b) Recorded negative phase contrast image of the phase distribution in (a). (c) Checkerboard phase distribution ($\psi_{CB}$) added to (a) using the SLM. (d) Acquiring a negative phase contrast image of $\psi=\psi_{CZ}+\psi_{CB}$ and subtracting it from the image in (b) reveals the intensity changes due to $\psi_{CB}$. in some areas of the sample  (e) Iterative phase imaging and wavefront correction leads to an artifact free image of $\psi_{CZ}$ (see appendix). A negative phase contrast image of $\psi=\psi_{CZ}-\psi_{meas}+\psi_{CB}$ now reveals the checkerboard with good signal-to noise across the field of view, demonstrating the sensitivity enhancement offered by LowPhi. All scalebars are 400 microns.}
\label{fig:Fig2}
\end{figure}

We tested our technique using a known sample. Using the SLM in a plane conjugate to the specimen plane, we can simply emulate a known phase distribution. Fig. \ref{fig:Fig2}a shows the phase distribution displayed on the SLM ($\psi_{CZ}$, Cezanne’s ‘Still Life with Apples’), spanning $2\pi$ radians. The corresponding recorded negative phase contrast image is shown in Fig. \ref{fig:Fig2}b (unity magnification). To test the local sensitivity to small phase-shifts, local changes in the form of a checkerboard ($\psi_{CB}(x,y)$, Fig. \ref{fig:Fig2}c) are added to the ‘Still Life with Apples’, again using the SLM, corresponding to changes in optical path length (OPL) of 10 nm for bright vs. dark squares. A negative phase contrast image of the summed phase distribution $\psi (x,y) = \psi_{CZ} (x,y)+\psi_{CB} (x,y)$  is taken and subtracted from the image in Fig. \ref{fig:Fig2}b. The difference is shown in Fig. \ref{fig:Fig2}d. While the changes introduced by the checkerboard are clearly seen in some regions of the image, the contrast vanishes in other regions, where sensitivity goes to zero. This clearly demonstrates that phase contrast microscopy is not quantitative and does not provide homogeneous sensitivity across the field of view - both as expected according to Eq. \ref{eq:eq1}.

When initializing LowPhi, the SLM is set to be a flat mirror. In this configuration, an estimate of the phase distribution can be readily obtained using a traditional quantitative phase imaging (QPI) technique \cite{Popescu2004}. The SLM is then used to subtract the measured phase-shifts from the wave-front. If another QPI image is acquired consecutively, it will show the errors of the initial estimate. These errors can be due to halo effects and other imaging artifacts, which are a common problem when imaging samples, in which phase shifts vary from $(-\pi, \pi)$ and show both large and small phase gradients. 
In order to properly initialize LowPhi (such that $\psi(x,y)-\psi_{meas}(x,y)\lesssim 0.5$ all across the field of view), it thus can be necessary to quantify these errors $\psi_{err}(x,y)$, and to update the  estimate of the phase distribution, which can then be subtracted from the wave-front using the SLM. If necessary, this can be repeated iteratively ($\psi_{meas,n}(x,y)=\psi_{meas,n-1}(x,y)+\psi_{err}(x,y)$). In our setup, 3 to 5 iterations were required until the initialization condition of LowPhi was met (see appendix).   
Fig. \ref{fig:Fig2}e shows the measured phase distribution after 5 iterations of QPI following wave-front subtraction. It shows excellent agreement with Fig. \ref{fig:Fig2}a and no imaging artifacts, apart from a very low frequency spatial gradient (see appendix), that is inevitable when using a finite-size phase plate to impart $\alpha$ onto the wave-front. 

Now that LowPhi is initialized, temporal changes can be followed quantitatively and at optimal sensitivity with every frame that is acquired. We demonstrate this by subtracting $\psi_{meas,n}(x,y)$ from the original wave-front, and by again adding the checkerboard phase distribution to it. The measured local intensity changes (the checkerboard) are now clearly visible in a negative phase contrast image (Fig. \ref{fig:Fig2}f), with good signal to noise and with a constant signal across the field of view. 

\subsection*{Analysis}

\begin{figure}[t]
\centering
\fbox{\includegraphics[width=84mm]{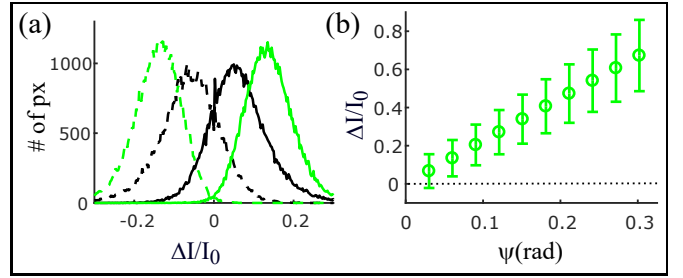}}
\caption{(a) Histogram of the relative intensity changes within the dark (dashed line) and bright (solid line) squares of the checkerboard, as measured via LowPhi (green) and phase contrast microscopy (black). (b) Distance of the peaks of the dashed and solid LowPhi histograms as a function of the phase value added by the checkerboard. The error bars indicate the widths of the histograms in (a).}
\label{fig:Fig3}
\end{figure}

This observation is quantified in Fig. \ref{fig:Fig3}: Fig. \ref{fig:Fig3}a shows a histogram of relative intensity changes due to the addition of the checkerboard phase distribution. The dashed (solid) lines give the results for the dark (bright) squares of the checkerboard. Black color indicates results obtained using phase contrast microscopy, green color indicates results from LowPhi.
The phase difference between the dark and bright squares is $\psi_{CB}=0.12$ rad, corresponding to a change of OPL of 10 nm. Before calculating the relative intensity changes, an unmodulated, incoherent background was subtracted from the images. It was measured by setting $\psi_{CB}$ to $\pi/2$ and fitting the intensity of the dark squares with a fifth order polynomial. Note that it represents an additional term in \eqref{eq:eq1} that does not affect the interference term. The mean signal, e.g. the distance between the peaks of the dashed and the solid line, is increased for LowPhi (green) as compared to negative phase contrast imaging (black), demonstrating the signal enhancing capabilities of LowPhi. Equally important is the reduction of the width of the two peaks ($\sigma_{bright}$ and $\sigma_{dark}$) for LowPhi, which leads to negligible overlap of the two peaks, and increased signal to noise. This is not the case for negative phase contrast imaging, where significant overlap does not allow for a quantitative phase measurement. Note, that while Fig.\ref{fig:Fig3}a demonstrates the global sensitivity enhancement obtained for this specific sample, the local sensitivity enhancements can be much larger, as evidenced by comparing Fig.\ref{fig:Fig2}d and Fig.\ref{fig:Fig2}f. 

LowPhi also yields a quantitative phase measurement. If the SLM is properly initialized, the deviations from a plane wave are small, $|E_S |\ll|E_R |$, and $|E_R |\approx|E |$. This yields $\Delta I(x,y) / I_0 \approx 2\psi(x,y)$, where $I_0=|E_R |^2$, as can be derived by rewriting and squaring Eq. \eqref{eq:eq0} as $|E_F |^2=||E_R |e^{i\alpha}+|E| e^{i\psi(x,y)}-|E_R ||^2$. Once properly initialized, LowPhi thus allows for single frame quantitative measurements, as long as $\Delta \psi(x,y) \lesssim 0.5$ rad during a time $\Delta t$, which characterizes the feedback loop of the system. Several times contribute to $\Delta t$: The exposure time, the time it takes to analyze a new image, and the inverse of the SLM refresh rate. Quantitative single frame results (after initialization) are demonstrated in Fig. \ref{fig:Fig3}b, where the measured relative intensity changes are plotted as a function of $\psi_{CB}$, varying from ~30 to 300 mrad (~2.5 to 25 nm in OPL).
For each value of  $\psi_{CB}$, histograms analogous to those in Fig. \ref{fig:Fig3}a are calculated. The error bars $\sigma$ in Fig. \ref{fig:Fig3}b denote the width of these histograms ($\sigma = \sqrt[]{\sigma_{bright}^2+\sigma_{dark}^2}$). 
For the smallest phase shifts the width is dominated by statistics as evidenced by a 1.97x reduction in width when averaging four images. Future experiments will deploy higher intensity light sources and higher sensitivity detectors to enable increased sensitivity and higher frame rates. While in the current study phase noise on the SLM, which is typically highest at frequencies >50 Hz, was averaged out, this will not be the case in high frame rate studies. It might then be necessary to synchronize the experiment to the SLM phase noise. 
For larger phase shifts, the width is currently dominated by spatial variations of the measured phase shifts. In future experiments, these can be reduced by a pixel-by-pixel calibration of the SLM and by a better estimate and reduction of the unmodulated background light. Note that the finite step size of present day spatial light modulators (typically 8bit) does not pose a limit to the technique, as discretization can be taken into account in image analysis.  

\subsection*{Imaging of Biological Specimens}
\begin{figure}[t]
\centering
\fbox{\includegraphics[width=84mm]{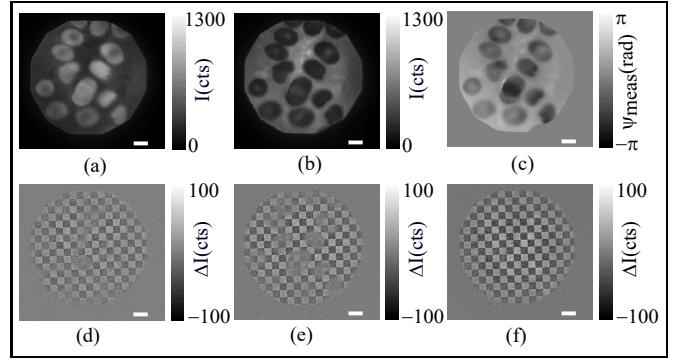}}
\caption{(a) Negative phase contrast, (b) positive phase contrast and (c) LowPhi images of red blood cells. Again, the local sensitivity to phase changes is probed by adding a checkerboard phase distribution, which is measured differentially as described in Fig.2 ((d), (e), and (f), respectively). All scalebars are 8 microns.}
\label{fig:Fig4}
\end{figure}

To test LowPhi under real imaging conditions, biological samples are introduced in the sample plane of the inverted phase microscope (see Fig. \ref{fig:Fig1}). A 50x objective is used to image them onto the SLM, which is used to emulate the small phase shifts (checkerboard), as well as for LowPhi. Fig. \ref{fig:Fig4}a,b, and c show the images of red blood cells obtained in negative phase contrast, positive phase contrast, and LowPhi, respectively. The right column shows the respective results of differential checker-board measurements, where $\psi_{CB}=0.12$ rad. For positive and negative phase contrast microscopy the local changes (the checkerboard) are only visible in some areas of the sample. Using LowPHi, the checkerboard is seen across the field of view, and with optimized sensitivity. Again, quantitative measurements can now be obtained from every single frame that is acquired.

\section*{Conclusion}
LowPhi is a technique that uses wave-front shaping to optimize, and thus homogenize, the sensitivity of phase imaging across the field of view. Once initialized, LowPhi allows for single frame quantitative phase imaging with minimal imaging artifacts such as halos. We demonstrated these key advantages of LowPhi with both artificial and biological samples and showed significant local sensitivity enhancement. LowPhi is a common path technique and thus very stable. It can be set-up as an add-on to a conventional (inverted) phase contrast microscope. We stress however, that the principle can also be applied to techniques using an external reference wave, in which either the reference or the signal wave can be reshaped to guarantee optimal sensitivity. 
Besides high speed quantitative studies of cell dynamics and morphology, LowPhi may also find use in multi-pass microscopy \cite{Juffmann2016,Klopfer2016} to avoid phase wrapping problems. To achieve this, an SLM would have to be incorporated within the self-imaging cavity of a multi-pass microscope, again in a plane that is conjugate to the sample. Considering recent advances in electron optics \cite{Verbeeck2017,Juffmann2017}, LowPhi might also be interesting for electron microscopy, where, due to specimen damage, it is absolutely crucial to achieve highest sensitivity per electron-sample interaction \cite{Glaeser2016a}.

\section*{Acknowledgments}
The authors thank J. Dong for helpful discussions, and S. Leedumrongwatthanakun and C. Knobloch for their help with the preparation of Figure 1.

\section*{Author contributions statement}
T.J. and S.G. conceived the technique, T.J. and A.S. performed the experiments,  T.J. and S.G. analyzed the results.  All authors reviewed the manuscript. 

\section*{Funding Information}
HFSP Cross-Disciplinary Fellowship (LT000345/2016-C); ERC SMARTIES (Grant 724473);
Institut Universitaire de France;

\section*{Additional information}
The author(s) declare no competing interests.

\appendix

\section{Folded LowPhi configuration}

\begin{figure}[htbp]
\centering
\fbox{\includegraphics[width=0.6\linewidth]{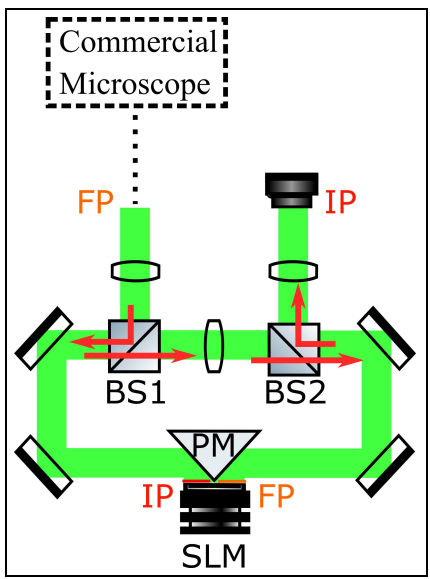}}
\caption{Folded LowPhi configuration (see text). }
\label{fig:folded}
\end{figure}
Figure \ref{fig:folded} shows the LowPhi configuration that was used in the experiment. A single reflective SLM is used to modulate the wave-front in a plane conjugate to the sample as well as in the Fourier plane. 

First, a wavefront coming from a commercial microscope is in-coupled using a beamsplitter (BS1) and imaged onto the left half of the SLM, which is conjugate to the sample. Then, a folded 2f system is used to propagate the wavefront to the right half of the SLM, where the wavefront can be modulated in Fourier space. A second beam splitter (BS2) is used to direct the remaining fraction of the wavefront to a CMOS detector.  
To minimize reflections back and forth between the two planes, BS1 was chosen to have a 90 to 10 ratio of reflected to transmitted intensity. A knife-edge right-angle prism mirror (PM) was used to address the two halves of the SLM at normal incidence.

\section{LowPhi initialization and image artifact removal}
\begin{figure}[htbp]
\centering
\fbox{\includegraphics[width=84mm]{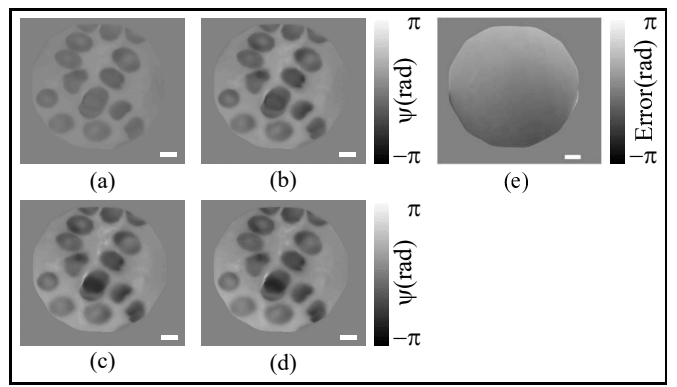}}
\caption{(a-d) Iterative initialization of LowPhi (see text). All scalebars are 8 microns. (e) artifact removal in LowPhi (see text). Scalebar is 400 microns.}
\label{fig:iterations}
\end{figure}

The initialization of LowPhi is exemplified with a series of four quantitative phase images (QPI) of red blood cells (Fig. \ref{fig:iterations}(a-d)). In our setup, the QPI were acquired using the SLIM routine \cite{Popescu2004}. The first such image underestimated the phase shifts that were present, which is due to the significant overlap of scattered light with the phase shifting element (PM) in the Fourier plane. The measured phaseshifts were subtracted from the wavefront using the SLM, and another quantitative phase image is acquired, which reveals the errors in the first image, and is used to update the SLM. This iterative initialization converges after three iterations, at which point all further phase changes are $ \ll 1$ rad. 

%\begin{figure}[htbp]
%\centering
%\fbox{\includegraphics[width=\linewidth]{FigS2}}
%\caption{artifact removal in LowPhi (see text). All scalebars are 400 microns. }
%\label{fig:artifacts}
%\end{figure}
Fig. \ref{fig:iterations}e shows the difference between the applied phaseshifts (see Fig.\ref{fig:Fig2}a) and the measured phaseshifts (see Fig. \ref{fig:Fig2}e) for the known phase distribution displaying Cezanne's 'Still Life with Apples'. It demonstrates that the algorithm converges to a result free of artifacts such as halos. Only a gradient remains, which corresponds to a spatial frequency too low to scatter light to regions outside the phase shifting element PM. This gradient could be minimized by reducing the radius of the PM when working with a light source of high transverse coherence \cite{Nguyen2017}.

\bibliographystyle{elsarticle-num} 
\bibliography{LowPhi}

\end{document}